# Robo Sapiens

Chaim Ash[1*], Amelia Hans[2*],

[1]RO-Folks'Talks ltd., Netanya, Israel; *chaim.ash@gmail.com

[2]Moona, Maj'd al-Krum, Israel; *ameliasimonehans114@gmail.com

**ABSTRACT**

This paper proposes a new method of natural language acquisition for robots that does not require the conversion of speech to text. Folks'Talks employs voice2voice technology that enables a robot to understand the meaning of what it is told, and to have the ability to learn and understand new languages - inclusive of accent, dialect, and physiological difference. To do this, sound processing and computer vision are incorporated to give the robot a sense of spatiotemporal causality. The 'language model' we are proposing equips a robot to imitate a natural speaker's conversational behavior by thinking contextually and articulating its surroundings.

## INTRODUCTION

Gary Marcus and Ernest Davis suggest that in order for machines to achieve true intelligence, we should draw inspiration from the structure of our own minds, especially those of young children who are capable of absorbing and comprehending new concepts far better than any machine, even though they cannot read.[1] From this stems our belief that text-based Natural Language Processing (NLP) is a dead end to understanding a language. Language acquisition is a step-by-step, multi-threaded process, that does not have any shortcuts.

Charles Sanders Peirce wrote: *We can see that a thing is blue or green, but the quality of being blue and the quality of being green are not things which we see*.[2] Peirce believed that quality is never solely an object of observation because it is a product of logical reflection; qualities are taught, and the learning of qualities in a particular way influences how people develop, perceive, interpret, communicate with, and understand their environment. Ultimately, this is how humans naturally form *guiding principles* which help to create a shared understanding of what is important, and which provide a framework for making decisions and taking actions that align with their values and goals. These guiding principles are not rigid rules, but rather flexible guidelines that can be adapted to changing circumstances and challenges.

---

[1] Marcus, Gary, and Ernest Davis (2019). Rebooting AI: Building Artificial Intelligence We Can Trust. Pantheon Books., p. 25.

[2] Peirce, Charles Sanders. "The Fixation of Belief." The Essential Peirce: Selected Philosophical Writings, Volume I (1867-1893), edited by Nathan Houser and Christian Kloesel, Indiana University Press, 1992, 109-123. Originally published in Popular Science Monthly, vol. 12 (November 1877), 1-15.

If we want to teach robots to understand us, then we should begin by teaching them both the physical and abstract qualities of objects, so that they may reflect, imagine, and *think* contextually about events - giving them a grasp of causality - the most crucial element in the creation of Strong AI.[3] With this capability, robots will be able to interact naturally with humans.

While we acknowledge that individuals who are deaf from birth can acquire language through sign language, this paper focuses on the development of technology that relies on spoken language. This is not intended to discount the importance of sign language as a valuable mode of communication.

## THE FOLKS'TALKS PROJECT: LANGUAGE ACQUISITION BY A VIRTUAL AGENT

We started this project as a computer simulator where a virtual agent acquires language from a native speaker. For the purpose of language acquisition, our research utilizes verbal language instead of written text. This is because spoken language is the primary way that people learn and use language, while writing is a secondary system.

**Primary Principles**

a. Acquisition starts from scratch. There is no preliminary data.
b. The data collected is vocal (in the form of wave files) and is not connected to any writing system.

---

[3] Pearl, Judea, and Dana Mackenzie (2018). The Book of Why: The New Science of Cause and Effect. Hachette UK.

c. Within the screen interaction areas of the simulation, graphical objects are placed in specific arrangements within specific scenes and with specific relationships. These objects represent the inner world of the simulation, which is a small reflection of the real world and its three main components: space, time, and society.

d. This inner world must be described vocally by a human speaker using phrases with distinct patterns. These phrases are the Vocal Markers (VMs) in the scene and include: (1) A phrase pattern; (2) an object within the phrase; (3) phrase intonation (questions, imperatives, statements, etc.); (4) number of words in the phrase; (5) the function of each word in the phrase; and (6) other markers such as reactions, intonations, and parts of speech

The speaker initially records and vocally marks a sufficient number of variations of the phrase patterns, with approximately five repetitions (for this test only) for each tested phrase. Then 53 sound features are extracted from each recording frame using the Essentia library. For subsequent object recognition and training of the virtual agent, the TensorFlow C-API is used. The trained virtual agent is then able to: (1) Name an object to which the human speaker points; (2) find a requested object; (3) give a suitable answer from a question-answer pair it is trained on; (4) give an answer regarding an object's size and color; (5) execute requested vocal commands; and (6) recognize different speaking intonations such as commands, questions, stories, etc.

These principles and their applications, which are demonstrated in 'Machine Learning for Speech Recognition' below, are suitable for the acquisition of any language.

**The Preliminary Study**

To determine the simplest initial training environment, a test was conducted with $6^{th}$ grade students (11–12-year-olds). The students were asked to describe a world comprised of only two objects (a block and a ball), each possessing a single property from a possible two colors and two sizes: green or red, and big or small. Following a series of tests, a language acquisition simulation model in the form of a computer game was selected for further development purposes.

The computer game was designed to provide a voice-based interaction between the player and a Virtual Agent (VA). The player and the virtual agent engage in question-answer and command-execution interaction. They also hold conversations regarding the virtual agent's navigation ('understanding') within the virtual world. *See Figure 3*.

The simplified virtual world is comprised of pushbuttons that symbolize real objects with their properties and interrelated qualities, such as relative size or position, as well as a display of the status of events through time representation.

The player records and tags phrase patterns related to the objects in the virtual world. The virtual agent is then trained to recognize these tagged phrases. Following the training, the virtual agent engages in interaction with the player through *the player's* natural conversation.

**Feature Extraction**

While interacting with a player, the voice recorder was always active and provided a list of wave files which are represented in a numerical form for subsequent feature extraction. Various libraries and toolkits, including Sphinx, Librosa, Kaldi, Pyaudioanalysis, and others were evaluated for this task. We found the Essentia library suitable for the aims of this project.

By using Essentia, we performed windowing (framing) and extracted spectrum and MFCC features for each frame, which provided us a 2D numerical raw array of size 53x (number of frames) which was converted to a normalized array of the same size for further use.

```cpp
essentia::standard::Algorithm* loader;
essentia::standard::Algorithm* fc;
essentia::standard::Algorithm* w;
essentia::standard::Algorithm* spec;
essentia::standard::Algorithm* mfcc;
loader = factory.create( "EasyLoader",
                         "filename",   audioFilename,
                         "sampleRate", analysisSampleRate,
                         "startTime",  startTime,
                         "endTime",    endTime,
                         "downmix",    downmix);
fc = factory.create(
                     "FrameCutter"
                     ,"frameSize", frameSize
                     ,"hopSize", hopSize
                     );
w    = factory.create("Windowing", "type", "square");
spec = factory.create("Spectrum");
mfcc = factory.create("MFCC");
```

All three charts within the following image depict the same Russian word 'Dva', which means 'two' in English. The red represents the normalized 2D array, while the black is the sound WAV file. The blue and orange sections represent the spectrogram of the sound file. The x-axis shows time, and the y-axis shows the extracted features of the word:

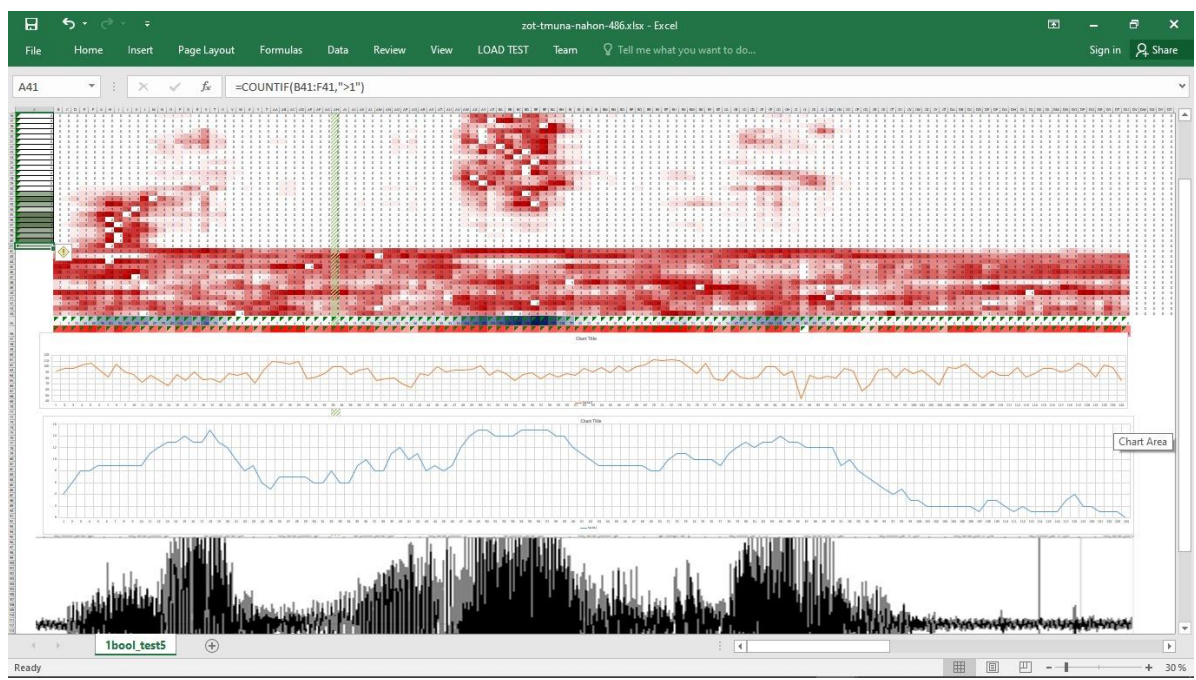

*Figure 1:* *Normalized 2D array, WAV file, and spectrogram of the sound file.*

**Machine Learning for Speech Recognition**

We selected the TensorFlow library for machine learning. To facilitate further study and improve work convenience, we created a graphical research interface. We evaluated various inputs and methods of normalization for the numerical arrays of sound features, and conducted preliminary tests in which we trained the virtual agent to recognize vocal counting in English, Russian, Hebrew, and Urdu. Next, we tested the virtual agent's ability to interact with a player within a virtual world consisting of five objects (our initial tests consisted of only two objects) and successfully conducted these tests in English, Russian, and Hebrew.[4]

The current development status includes a virtual world with ten objects described using 22 phrase patterns, which include the object's name, a name-question-answer-command chain, and a name-property positive/negative-question-answer-command chain.

In the name-question-answer-command chain, we first give an object a name such as 'apple.' We then ask, 'What is it?' and provide the answer, 'This is an apple.' In the name-property positive/negative-question-answer-command chain, we start by stating 'This is an apple' and then ask, 'Where is the apple?', followed by the commands 'Show me an apple' and 'Give me an apple.' Once these facts are established, we can ask more sophisticated questions such as 'Where is the red apple?' The positive/negative component is determined by questions like 'Is it a green apple?' to which the answer is either 'Yes' or 'No.' This is a communication style commonly used by parents when speaking to their children.

**Database Creation**

The database consists of:

(1) A group of four text files – a data array for training, a data array for testing, a labels array for training, and a labels array for testing – placed in separate folders designated for objects, phrase patterns, intonation, and emotion. *See Figure 2*. There can be a greater number of files for any other feature we wish to explore.

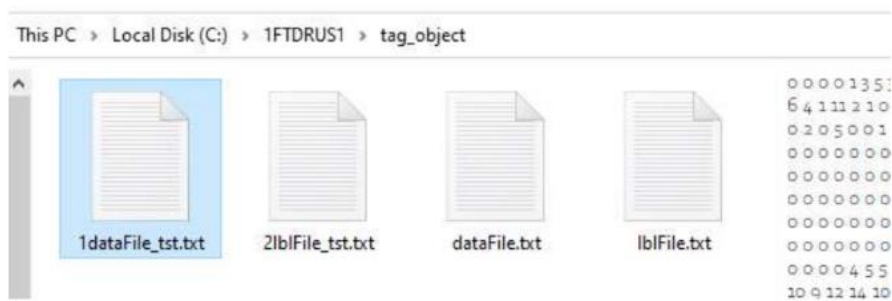

---

[4] https://youtu.be/DNh_tQ1mL6s ("Folks'Talks English test"); https://youtu.be/j33FwLQjwTI ("The Folks'Talks video game Hebrew test"); https://www.youtube.com/watch?v=3_IZ2_fjg5g ("Folks'Talks Final Russian Test"); https://www.youtube.com/watch?v=RyaBdgiHUlY ("Folks'Talks' translation ability test").

*Figure 2: This is a folder designated for storing the names of objects, with each folder containing exactly four files.*

(2) A collection of audio files saved after tagging.

(3) A text file (the 2d normalized array) of tagged lines for creating relevant combo-boxes in the Graphical User Interface (GUI). In this study, the text file was used by the virtual agent for asking questions and answering them.

Each group of four text files was converted to the NPZ format using the NumPy Python library. There is one NPZ file for each parameter – to be recognized by the virtual agent later. Each of the NPZ files was then converted by the TensorFlow library to Protocol Buffers (PB files), one for each recognized parameter.

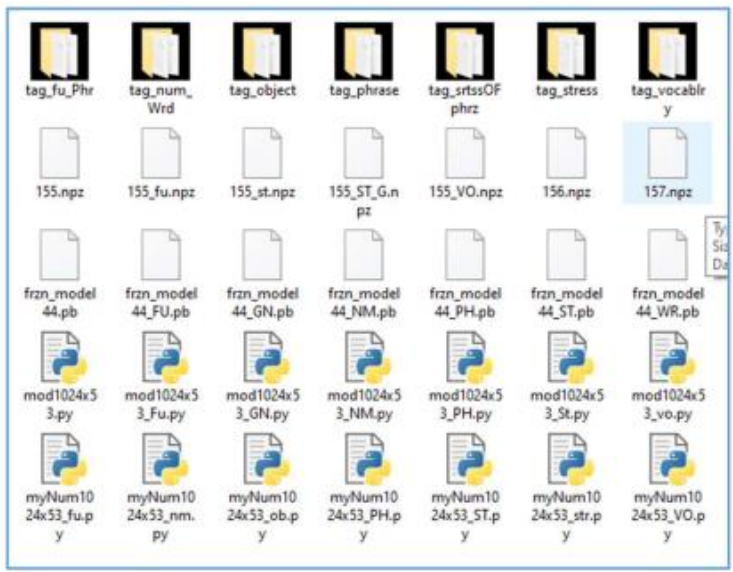

**Language Model**

We did not use a language model. Our virtual agent simply imitates the conversational behavior of a natural speaker. This simplified conversational behavior consists of: (1) Spotting an object; (2) asking what it is; (3) answering that the spotted object is a certain thing; (4) inquiring about the properties of an object; and then (5) executing commands or actions related to the spotted objects. All of these are common principles of language acquisition in children.

In order to simulate conversational behavior, we placed objects (such as a table, window, etc.) in a virtual scene and a native speaker described the scene using typical phrases in their mother tongue, as previously described using the example of an apple. We also asked the native speaker to make VMs for the spoken phrases (such as object ID, phrase pattern ID, phrase intonation, phrase separation into words, marking each separate word in the phrase).[5]

Word functions used for marking are:

- Person
- Alive
- Object
- Action
- Property name
- Object's property
- Action's property
- Functor (function word)
- Positive
- Negative
- Pointer

**Virtual Agent Training and Training Modes**

There are two modes in this project: 'Training Mode' and 'Talking Mode'.

**(1) Training Mode**

In Training Mode, a player maps and creates a database by delivering information to the computer using verbal language.

---

[5] Ibid.

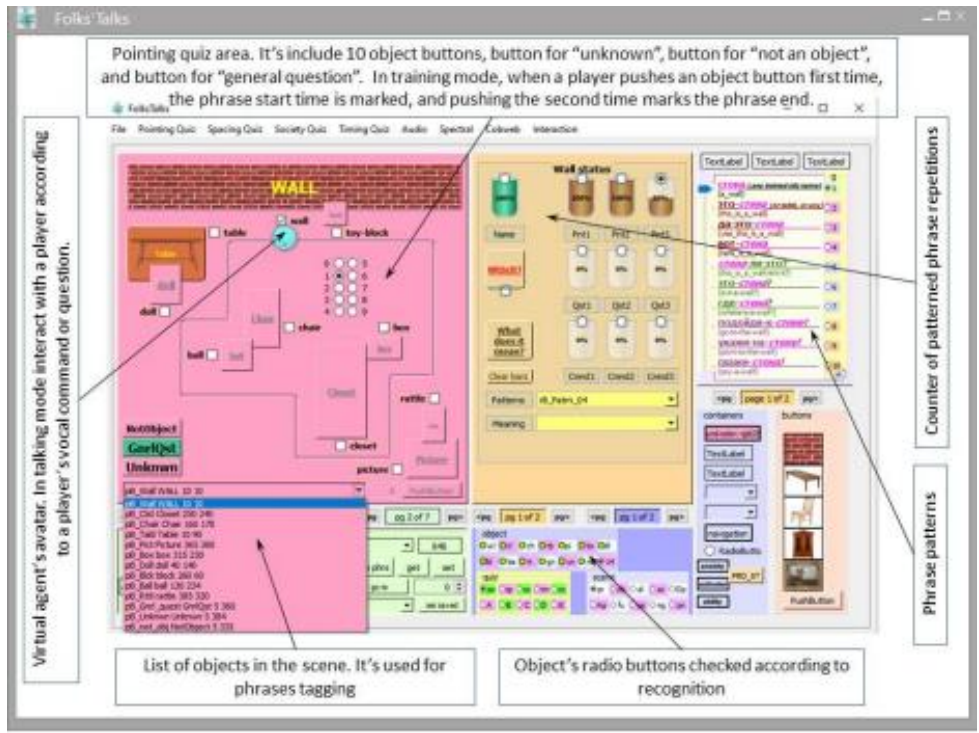

*Figure 3:* *There are several objects located around the ball. For purposes of this study, we conducted five repetitions. Once five repetitions are complete, the machine learning algorithm can be activated as there is enough data to train it. Similar to teaching children, repetition is necessary.*

This database will later serve for voice recognition, and the sound files from the database will be used for the virtual agent's voice response.

**(2) Talking Mode (VA-Player Interaction)**

For the purpose of our preliminary research, a simplified interaction method was implemented. There are two methods of delivering a message to the virtual agent: (1) By pressing the button that represents an object when the message requires physical identification ('pointing'); and (2) by pressing a general input button when an object's name is mentioned in a phrase.

The player's voice message is converted into a normalized array, and its parameters are recognized using the TensorFlow C-API:

```cpp
class my_tensor6
{
public:
    my_tensor6();
    static TF_Buffer *read_tf_buffer_from_file (const char* file);

 void load_BoolNew53XZ_Ob (unsigned char **); /*object*/
 void load_BoolNew53XZ_Ph (unsigned char **); /*phrase patterns*/
 void load_BoolNew53XZ_Nm (unsigned char **); /*number of words in phrase*/
 void load_BoolNew53XZ_St (unsigned char **); /*intonation of word in phrase*/
 void load_BoolNew53XZ_Gn (unsigned char **); /*general phrase intonation */
 void load_BoolNew53XZ_Fu (unsigned char **); /*word's function in phrase*/
 void load_BoolNew53XZ_Wr (unsigned char **); /*all words in VA vocabulary*/

 float my_pred_wrd [13]; /*objects*/
 float my_pred_phrz [22]; /*phrase patterns*/
 float my_pred_count [10]; /*number of words in phrase*/
 float my_pred_srts [4]; /*intonation of specific word in phrase*/
 float my_pred_gnStr [4]; /*general phrase intonation */
 float my_pred_func [11]; /*word's function in phrase*/
 float my_pred_vcab [40]; /*all words in agent's vocabulary*/
};
```

VA responses are based on recognized patterns and object names. The agent can either answer a question by retrieving a suitable phrase from its training mode or execute a command from the message, such as navigating to a relevant object.

If the VA makes a mistake, the player can mark the correct tags, and this correction will be incorporated into the next training session.

**Conclusion of Preliminary Study**

While we were successful in training the VA to recognize specific objects and understand what objects we referred to when speaking with it[6], this limited environment was insufficient for the VA to gain any true understanding of the real world. As noted by Marcus and Davis, ...*labeling the objects in a scene is less than half the battle. The real challenge of situational awareness is to understand what all those objects collectively mean.*[7] In other words, the open-endedness of the world cannot be taught through our simulation since the VA is confined within a 'game' and is unable to interact with real objects. To bridge this gap, a robot is needed. With a robot, learning language will no longer be limited by a two-dimensional setting, because the VA will have the ability to 'experience' the real world.

---

[6] Ibid.

[7] Marcus, op. cit., p. 106.

**MOVING FORWARD**

Since each stage of language acquisition requires a different approach to understanding, it is essential to address every stage if we want to create real human-robot interaction and achieve high-quality communication between them.

This is the order in which we propose to teach robots how to acquire language:

(1) **Object Name.** The initial stage of language acquisition involves gathering visual and geometrical information about objects. Robots equipped with 3D sensors and cameras can obtain meshes and pictures of nearby objects, including their position relative to defined reference points, which constitute an internal coordinate system, rather than just using a SLAM algorithm. This enables us to obtain a collection of photos and a 3D representation of each object, defined by its ID and coordinates within a given space.

We propose organizing these photos into a GUI and presenting them individually with accompanying sound or speech labels. These sound labels can be used later for question and answer pairs or for searching and pointing missions.

Similar to what we previously described with the apple, first a question such as 'What is it?' must be asked. Then we must say, 'This is an apple.' Next, we must ask, 'Is it an apple?' If the object we are pointing at is what we previously referred to as an apple, then the answer will be 'Yes, this is an apple'. Then we can ask, 'Where is the apple?' and the robot will answer, 'Here is the apple.' Although the robot may still be unable to describe the position of the apple, it can point to it.

(2) **Object Color.** Usually the name of an object's property goes together with the name of the object that has that property. The challenge is to differentiate the sound of the unknown property from the sound of the known object. The name of the property has no strict definition and can vary under the same sound. For example, 'green' can be green, light-green, dark-green, greenish, blue-green, etc.

We propose using toy blocks of various colors to visually represent the color properties that can later be marked with a vocal label. After that, language patterns like 'green apple' or

'yellow apple' could be used and distinguished on-site in real time. For example, after teaching the robot 'green', the user should show it a green apple and ask it: 'What is the color of the apple?', then: 'Where is the green apple?'

**(3) Size (space property).** In order to describe the size of an object, it's necessary to compare it to another object. This allows for a conclusion to be drawn regarding their relative sizes. This enables the robot to understand and answer questions such as: 'Where is the big/small apple?'

**(4) Commanding the robot to interact with a user (robot-object-user direction).** The intonation used to issue a command itself is enough to signal the need for action. The robot must then plan a trajectory of movements required to manipulate objects and complete the requested action. This requires creating a sequence of imaginary pictures depicting future situations, starting from the current position and ending with the task completion. During training, successful execution of the command is marked with a sign of approval. This enables the robot to understand and execute commands such as: 'Bring me an apple' and 'Help me find an apple.'

**(5) Commanding the robot to act in space (robot-object1-object2 direction).** This involves understanding prepositions and placing the requested objects in their relative positions. 'Put an apple in/on the fridge/counter.'

**(6) Society (ownership).** Understanding social and ownership relations. 'Give *him* the apple', 'That is *your* apple.'

**(7) Society (placeholder).** Understanding who he/she/they (anyone that is not the user) is. Being able to associate nicknames such as 'darling' and 'honey' with a particular person. Comprehending social (family) relations such as 'sister' and 'father'.

**(8) Society (direction of action).** '*Bring* him/her an apple'.

**(9) Events (collision).** This process involves reconstructing the possible beginning of an event by starting from the end when something has already happened, such as a change in form, position, condition, or property. It begins with defining the frames, including the event's start and end points/borders.

(10) **Event (imaginary).** Some events are visible from a specific point of view only, such as watching a sunrise from a particular location.

(11) **Event (repeated).** Re-imagining a sunrise.

(12) **Causality.** Something has fallen from the table. The robot has to able to understand why it fell. It must be able to reflect and conclude where the starting point of the event that led to this fall was.

(13) **Tenses.** At some point, children start to understand time. The robot must be able to do this as well and be able to respond accordingly when hearing words such as 'today,' 'tomorrow,' and 'yesterday.'

(14) **Emotions/feeling imitations.** A measurement alone cannot estimate an event or condition. A robot needs to be equipped with a tool that enables it to have a human-like indication of what is comfortable or not. For instance, 36ºC is cold for coffee but warm for ice cream. This can be taught to the robot using a slider for emotion level on the user interface.

(15) **Extract sentences from hearing.** Enables a robot to ask: 'What does this word mean?' or 'Can you clarify what you just said?'

(16) **Recollection/Reconstruction.** Initially there may be no information available about a possible future event. Robots need to receive a human estimation about the occurrence of a certain condition and then work backwards to reconstruct the event and 'propose' the critical point that may have led to that particular 'final' or 'resulting' condition.

(17) **Plans/fantasies.** After recognizing the 'command intonation,' the robot has to look ahead and plan (or recollect similar events from the past) all the future tasks it has been asked to perform. When the robot is in working mode, after a training phase based on reconstructions, the robot must anticipate or plan the future.

**CONCLUSION**

In this paper, we have presented a novel approach for a robot to acquire a true understanding of language by making every spoken sound and word correspond to specific objects, their appearance, and their behavior. This truly intelligent robot will be able to act according to a particular situation and will be capable of understanding its user's needs.

Over the past two years, more than 100 million assistive robots have been sold globally (according to Statista), but they lack the capacity to comprehend their environment and provide physical assistance, such as fetching glasses or other items. Our software addresses this issue, because it enables users to educate robots to recognize spoken language and assist in a range of scenarios, from home to medical care. This novel type of robot will not only be able to communicate, but also interact naturally and assist without the need for the user to view a display or take further steps.

While there are still some challenges that need to be addressed, such as preventing someone from teaching their robot to perform purposefully harmful actions, we believe that our work represents a promising step in the advancement of artificial intelligence that will lead us into an era of 'robo sapiens'.

**Acknowledgments**

My friends who patiently heard my fantasies and support me in the current along this project.

Amelia Hans was the primary author of this article but did not participate in the preliminary study. Raquel Lazar-Paley assisted with the final edits.

Chaim's grandchildren, Hadas (6 years old) and Lavi (4 years old) have been a source of inspiration and motivation for his work in robotics. Their curiosity, enthusiasm, and eagerness to learn have helped Chaim approach the challenge of language acquisition in robotics with a fresh perspective, encouraging him to explore new ideas and possibilities.